\documentclass[twocolumn,aps,showpacs,showkeys,floatfix]{revtex4}
\usepackage{epsfig,amsmath,amssymb}
\usepackage{amsmath}
\usepackage{color}

\begin{document}

\title{Exact solution for quantum spin-1/2 Ising-Heisenberg orthogonal-dimer chain
with the Heisenberg intra-dimer and Ising inter-dimer interactions}
\author{Taras Verkholyak$^{1,2}$,
        Jozef Stre\v{c}ka$^{2}$}
\affiliation{$^1$Institute for Condensed Matter Physics,
             National Academy of Sciences of Ukraine,
             1 Svientsitskii Street, L'viv-11, 79011, Ukraine\\
             $^2$Department of Theoretical Physics and Astrophysics,
             Institute of Physics, P. J. \v{S}af\'{a}rik University,
             Park Angelinum 9, 040 01 Ko\v{s}ice, Slovak Republic}
\date{\today}

\begin{abstract}
The quantum spin-1/2 orthogonal-dimer chain with the Heisenberg intra-dimer and Ising inter-dimer interactions
in a magnetic field is considered by a rigorous approach. The model conserves the $z$-component of total spin
on vertical Heisenberg bonds and this property is used to calculate exactly the partition function using the transfer-matrix method.
We have found the ground-state phase diagram of the given model in a magnetic field as well as
the macroscopic degeneracy along field-induced transitions accompanied with the magnetization jumps.
The model exhibits two intermediate fractional plateaux at one-quarter and one-half of the saturation magnetization.
We have examined the effect of the exchange anisotropy in the $XXZ$ Heisenberg intra-dimer interaction on the ground state.
It is shown that the one-quarter and one-half plateaux may disappear from the magnetization curve for
the ferromagnetic Heisenberg intra-dimer interaction. We have also studied rigorously the effect of frustrated interactions
on the thermodynamic and magnetic properties of the model and show how the macroscopic degeneracy of the ground state is
reflected in the low-temperature behavior of the magnetization, entropy and specific heat. A possibility of observing enhanced magnetocaloric
effect during the adiabatic demagnetization is discussed in detail.
\end{abstract}



\maketitle

%\clearpage

%\renewcommand\baselinestretch{1.15}
%\large\normalsize

\section{Introduction}

The quantum spin-1/2 dimer-plaquette or orthogonal-dimer chain \cite{richter1998,koga2000}
represents one of the known examples of partially exactly solvable models
with the dimerized ground state \cite{miyahara2011}.
Originally it was suggested as a one-dimensional counterpart of
the depleted square lattice \cite{richter1998} or the Shastry-Sutherland lattice \cite{miyahara2005}.
Despite its specific structure the Shastry-Sutherland model is related to a number of
magnetic compounds (SrCu$_2$(BO$_3$)$_2$, TmB$_4$, TbB$_4$, etc.) having either almost isotropic Heisenberg
or highly anisotropic Ising interactions (see Ref. \cite{takigawa2011} for a recent review).
Magnetization curves of these compounds exhibit the set of fractional plateaux,
which has not been firmly explained yet  \cite{takigawa2011}.
A consistent explanation of the series of plateaux in a low-temperature magnetization curve of
SrCu$_2$(BO$_3$)$_2$ remains the hot topic of the current research \cite{takigawa2013,matsuda2013}.
The comprehensive study of the model is in general quite difficult
and the only known exact result concerns the dimerized ground state for the sufficiently strong
intra-dimer (diagonal) interactions \cite{shastry1981}.
Since the orthogonal-dimer chain or two coupled orthogonal-dimer chains
\cite{manmana2011} can approximate to some extent the Shastry-Sutherland lattice,
their study may reveal some basic features which are typical also for this frustrated two-dimensional model.
For instance, the quantum spin-1/2 Heisenberg orthogonal-dimer chain shows
an infinite series of the magnetization plateaux \cite{schulenburg2002,schulenburg2002a}.
At the same time the Heisenberg model of two coupled orthogonal-dimer chains considered in Ref.~\cite{manmana2011}
exhibits a number of fractional plateaux and some of them are identical to the ones observed
in the Shastry-Sutherland model.

Recently, a quite specific quantum spin-1/2 orthogonal-dimer chain with triangular $XXZ$ Heisenberg clusters coupled
via the intermediate Ising spins has been considered by Ohanyan and  Honecker \cite{ohanyan2012}.
This simplified Ising-Heisenberg orthogonal-dimer chain is exactly soluble by means of the transfer-matrix method
and shows under certain conditions surprisingly good correspondence to the pure quantum model with all Heisenberg interactions.
In the present work, we will study another version of the spin-1/2 Ising-Heisenberg orthogonal-dimer chain, where the quantum Heisenberg
interactions are retained on all vertical and horizontal bonds coupled together through the Ising interactions.
This model preserves the $z$-component of the total spin on vertical Heisenberg bonds (dimers), which allows us
to obtain exactly all ground states and thermodynamic properties. During the preparation of this work, we became aware
of the similar work treating the same spin-1/2 Ising-Heisenberg orthogonal-dimer chain in an absence of the external field
using somewhat different approach based on a direct algebraic mapping transformation \cite{paulinelli2013}.

Last but not least, let us provide some insight into an experimental background of our work. Although the exactly solved Ising-Heisenberg models
with alternating Ising and Heisenberg bonds could be regarded more as a mathematical curiosity rather than
the realistic models of some  actual
magnetic materials, recent progress in the field of magnetochemistry has opened up new possibilities for a targeted design of magnetic materials
with a very specific combination of magnetic interactions. A few eminent Ising-Heisenberg models have proved their usefulness by an explanation of the magnetic
behavior of some real insulating magnetic materials such as [(CuL)$_2$Dy][Mo(CN)$_8$] \cite{heu10}, [Fe(H$_2$O)(L)][Nb(CN)$_8$][Fe(L)] \cite{sah12}
and Dy(NO$_3$)(DMSO)$_2$Cu(opba)(DMSO)$_2$ \cite{str12,han13}.

\begin{figure}[t]
 \begin{center}
   \epsfig{file=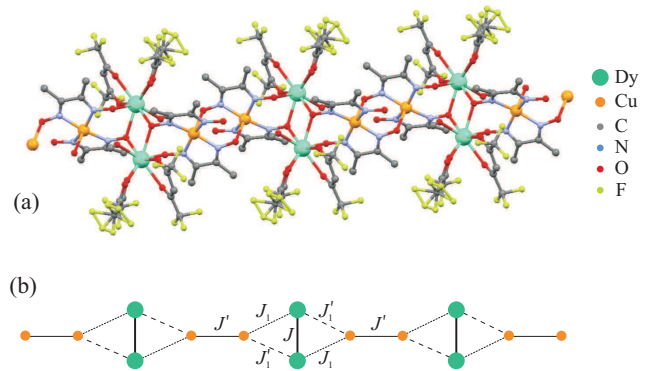, width=1\columnwidth}
 \end{center}
\caption{(Color online) (a) A segment from the crystal structure of the bimetallic polymeric chain Dy(hfac)$_2$(CH$_3$OH)]$_2$[Cu(dmg)(Hdmg)]$_2$ ([Dy$_2$Cu$_2$]$_n$) adapted according to the crystalographic data reported in Ref. \cite{okazawa2008}. For better clarity, the crystalographic positions of hydrogen atoms are omitted and two metallic atoms (Dy and Cu) are shown by the balls with two times larger van der Waals radii than the balls representing the non-metallic atoms; (b) A schematic representation of the magnetic structure of [Dy$_2$Cu$_2$]$_n$, which corresponds to the spin-1/2 orthogonal-dimer chain with four different exchange pathways between Dy-Dy, Cu-Cu and Dy-Cu magnetic ions.}
\label{fig_Dy-Cu}
\end{figure}

A series of isostructural 3d-4f coordination polymers [Ln(hfac)$_2$(CH$_3$OH)]$_2$[Cu(dmg)(Hdmg)]$_2$ (Ln = Gd, Dy, Tb, Ho, Er, Pr, Nd, Sm, Eu) involves an unprecedented heterobimetallic motif of the orthogonal-dimer chain \cite{ueki2005,ueki2007,okazawa2008,okazawa2009,okazawa2011}. The dysprosium-based member of this isomorphous series Dy(hfac)$_2$(CH$_3$OH)]$_2$[Cu(dmg)(Hdmg)]$_2$ to be further abbreviated as [Dy$_2$Cu$_2$]$_n$ (see Fig.\ref{fig_Dy-Cu}(a)) provides a valuable experimental realization of the spin-1/2 Ising-Heisenberg orthogonal-dimer chain due to a rather strong magnetic anisotropy of Dy$^{3+}$ ions \cite{ueki2007,okazawa2008}. As a matter of fact, the vertical spin-1/2 Ising dimers assigned to double oxo-bridged dinuclear entities of Dy$^{3+}$ ions regularly alternate within the polymeric compound [Dy$_2$Cu$_2$]$_n$ with the horizontal spin-1/2 Heisenberg dimers assigned to the macrocyclic dinuclear entities of Cu$^{2+}$ ions. It turns out that the antiferomagnetic superexchange coupling between Dy$^{3+}$ and Cu$^{2+}$ ions mediated by the oximate bridge is by far the most dominant coupling, whereas the superexchange mechanism for the double oxo-bridged dinuclear entities of Dy$^{3+}$ ions and the macrocyclic dinuclear entities of Cu$^{2+}$ ions transmit presumably much weaker ferromagnetic coupling \cite{okazawa2008}.
It can be clearly seen from Fig.\ref{fig_Dy-Cu}(a) that the magnetic structure of [Dy$_2$Cu$_2$]$_n$ implies a more complex (asymmetric) interactions between the spin-1/2 Ising and Heisenberg  dimers, which involve in total four different exchange pathways between Dy-Dy, Cu-Cu and Dy-Cu magnetic ions (see Fig.\ref{fig_Dy-Cu}(b)). For the sake of simplicity, we will restrict our further analysis only to the symmetric particular case with just two different exchange couplings (see Fig.~\ref{od-chain}), whereas the comprehensive analysis of the more general (asymmetric) case involving four different exchange couplings will be the subject matter of our future work.

The paper is organized as follows. In Section \ref{model_section} we introduce the spin-1/2 Ising-Heisenberg orthogonal-dimer chain
with the alternating Heisenberg and Ising interactions and solve it using the transfer-matrix method.
In Section \ref{gs_section} we will consider in detail the ground-state phase diagram.
Section \ref{thermodynamics_section} presents the most interesting results for the thermodynamic quantities and
the magnetocaloric effect. The most important findings are briefly summarized in Section \ref{conclusions}.

\section{Model and solution}
\label{model_section}

Let us define the quantum spin-1/2 Ising-Heisenberg orthogonal-dimer chain with the Heisenberg intra-dimer and the Ising inter-dimer interactions
in a magnetic field through the Hamiltonian (see Fig.~\ref{od-chain}):
\begin{figure}[t]
 \begin{center}
   \epsfig{file=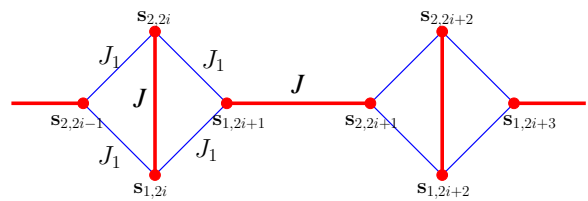, width=0.9\columnwidth}
 \end{center}
\caption{(Color online) The spin-$\frac{1}{2}$ Ising-Heisenberg orthogonal-dimer chain.
Thick (thin) lines denote the Heisenberg (Ising) bonds.}
\label{od-chain}
\end{figure}
\begin{eqnarray}
\label{gen_ham1}
H&{=}&\sum_{i=1}^N H_i,
\\
H_{2i+1}&{=}&
J_1[(s_{1,2i}^z{+}s_{2,2i}^z)s_{1,2i+1}^z{+}s_{2,2i+1}^z(s_{1,2i+2}^z{+}s_{2,2i+2}^z)]
\nonumber\\&&
+J({\mathbf s}_{1,2i+1}\cdot{\mathbf s}_{2,2i+1})_{\Delta}
{-}h(s_{1,2i+1}^z{+}s_{2,2i+1}^z),
\nonumber\\
H_{2i}&{=}&
J({\mathbf s}_{1,2i}\cdot{\mathbf s}_{2,2i})_{\Delta}-h(s_{1,2i}^z+s_{2,2i}^z),
\nonumber
\end{eqnarray}
where
$({\mathbf s}_{1,i}\cdot{\mathbf s}_{2,i})_{\Delta}
=s_{1,i}^x s_{2,i}^x + s_{1,i}^y s_{2,i}^y + \Delta s_{1,i}^z s_{2,i}^z$,
$s_{l,i}^\alpha$ denotes spatial projections ($\alpha=x,y,z$) of the spin-$\frac{1}{2}$ operator,
$J$ is the anisotropic Heisenberg intra-dimer interaction between spins on vertical and horizontal bonds,
$\Delta$ is the anisotropy parameter and $J_1$ is the Ising inter-dimer interaction between spins from different bonds.
In what follows, we will be mainly interested in investigating the particular case of antiferromagnetic interactions $J>0$, $\Delta>0$, $J_1>0$,
which brings the spin frustration into play.
Further, the periodic boundary condition for spins ${\mathbf s}_{l,N+1}\equiv{\mathbf s}_{l,1}$
will be implied for convenience.

Since $z$-component of the total spin on a vertical Heisenberg bond is the integral of motion,
and it is the only common operator for neighboring local Hamiltonians $H_i$,
all $H_i$ commute with each other. Hence, it follows that it is quite convenient to use a decomposition
of the total Hamiltonian (\ref{gen_ham1}) into the sum of commuting parts $H=\sum_{i=1}^{N/2}\tilde{H}_{2i+1}$, where
\begin{eqnarray}
\label{local_ham}
\tilde{H}_{2i+1}=H_{2i+1}+(H_{2i}+H_{2i+2})/2.
\end{eqnarray}
Consider now the total spin momentum operator on the vertical Heisenberg bonds ${\mathbf S}_{2i}={\mathbf s}_{1,2i}+{\mathbf s}_{1,2i}$.
It is quite apparent that ${\mathbf S}_{2i}$ represents the conserved quantity with well defined quantum numbers $S=0,1$
and $|{\mathbf S}_{2i}|^2=S(S+1)$, $S_{2i}^z=-S,-S+1,\dots,S$ \cite{rojas2012}. The respective eigenstates of this momentum spin operator
can be denoted as $|S_{2i},S_{2i}^z\rangle$.
%One can easily get
%$({\mathbf s}_{1,2i}{\mathbf s}_{2,2i})_{\Delta}
%=\frac{1}{2}(|S_{2i}|^2+(\Delta-1)|S_{2i}^z|^2)-\frac{1}{4}J(2+\Delta)$,
%$s_{1,2i}^z s_{2,2i}^z=\frac{1}{2}|S_{2i}^z|^2-\frac{1}{4}$.

Using the transfer-matrix method \cite{baxter}, the partition function of the model can be written in the form:
\begin{eqnarray}
Z
%=\mbox{Tr}\exp(-\beta H)
=\mbox{Tr}_{\{S_{2i},S_{2i}^z\}}\prod_{i=1}^{N/2}T(S_{2i},S_{2i}^z;S_{2i+2},S_{2i+2}^z),
\end{eqnarray}
where the transfer-matrix
$T(S_{2i},S_{2i}^z;S_{2i+2},S_{2i+2}^z)=\mbox{Tr}_{\{s_{1,2i+1},s_{2,2i+1}\}}\exp(-\beta \tilde{H}_{2i+1})$
contains the trace over two spins from the $(2i+1)$-st horizontal Heisenberg bond. Here, $\beta=1/T$ denotes the inverse temperature (Boltzmann's constant
is set to unity $k_{\rm B} = 1$). The straightforward calculation gives the transfer-matrix in the form
(where rows and columns  corresponds to the following set of states
$|0,0\rangle$, $|1,1\rangle$, $|1,0\rangle$, $|1,-1\rangle$):
\begin{eqnarray}
\label{t-matrix}
&&T=\left(
\begin{array}{cccc}
a_4 b_2^2     & a^{-}_3 b^{-}_1 b_2    & a_4 b_2^2 c     & a^{+}_3 b^{+}_1 b_2 \\
a^{-}_3 b^{-}_1 b_2   & a^{-}_1 (b^{-}_1)^2      & a^{-}_3 b^{-}_1 b_2 c   & a_2 b^{-}_1 b^{+}_1 \\
a_4 b_2^2 c   & a^{-}_3 b^{-}_1 b_2 c  & a_4 b_2^2 c^2   & a^{+}_3 b^{+}_1 b_2 c\\
a^{+}_3 b^{+}_1 b_2 & a^{-}_1 b^{-}_1 b^{+}_1   & a^{+}_3 b^{+}_1 b_2 c & a^{+}_1 (b^{+}_1)^2
\end{array}
\right),
%\\
\end{eqnarray}
\begin{eqnarray}
\label{coeff}
&&a^{\pm}_1{=}2\left\{ \mbox{e}^{{-}\frac{\beta\Delta J}{4}} \cosh[\beta(J_1{\pm}h)]
     {+}\mbox{e}^{\frac{\beta\Delta J}{4}} \cosh\left(\frac{\beta J}{2}\right) \right\},
\nonumber\\
&&a_2{=}2\left\{ \mbox{e}^{{-}\frac{\beta\Delta J}{4}} \cosh(\beta h)
     {+}\mbox{e}^{\frac{\beta\Delta J}{4}} \cosh\left( \frac{\beta }{2}\sqrt{J^2{+}4J_1^2}\right) \right\},
\nonumber\\
&&a^{\pm}_3{=}2\left\{ \mbox{e}^{{-}\frac{\beta\Delta J}{4}} \cosh\left[\beta \left(\frac{J_1}{2}{\pm}h\right)\right]
\right.
\nonumber\\
     &&\left.{+}\mbox{e}^{\frac{\beta\Delta J}{4}} \cosh\left( \frac{\beta }{2}\sqrt{J^2{+}J_1^2}\right) \right\},
\nonumber\\
&&a_4{=}2\left\{ \mbox{e}^{{-}\frac{\beta\Delta J}{4}} \cosh(\beta h)
     {+}\mbox{e}^{\frac{\beta\Delta J}{4}} \cosh\left( \frac{\beta J}{2}\right) \right\},
\nonumber\\
&&b^{\pm}_1{=}\mbox{e}^{{-}\frac{\beta}{2}\left(\frac{\Delta J}{4}{\pm}h\right)},\:
b_2{=}\mbox{e}^{\frac{\beta J(2{+}\Delta)}{8}}, \:
c{=}\mbox{e}^{{-}\frac{\beta J}{2}}.
\nonumber
\end{eqnarray}
Since the first and third row of the transfer matrix (\ref{t-matrix}) are linearly dependent, the transfer-matrix $T$ is the degenerate matrix, and at least one of the eigenvalues equals zero. In the case of zero external field  $a^{\pm}_1=a_1$, $a^{\pm}_3=a_3$, $b^{\pm}_1=b_1$ and consequently, it is possible to find the eigenvalues of the transfer matrix as two simple roots and two roots of quadratic equation \cite{paulinelli2013}.
%\begin{eqnarray}
%\label{ev1}
%lambda_1&=&0,
%\nonumber\\
%\lambda_2&=&(a_1-a_2)b_1^2,
%\nonumber\\
%\lambda_{3,4}&=&\frac{1}{2}(a_4 b_2^2(1{+}c^2){+}(a_1{+}a_2)b_1^2  )
%\nonumber\\&&
%{\mp}\frac{1}{2}
%\sqrt{(a_4 b_2^2(1{+}c^2){-}(a_1{+}a_2)b_1^2)^2 {+} 8a_3^2 b_1^2 b_2^2(1{+}c^2)}.
%\end{eqnarray}
%It is readily seen that $\lambda_4=\lambda_{max}$. At first it is positive $\lambda_4>\lambda_1$,
%since all the coefficients in the expression is positive,
%and $\lambda_4>\lambda_3$. Further we can see $\lambda_2<(a_1+a_2)b_1^2$.
%If $(a_4 b_2^2(1{+}c^2)<(a_1{+}a_2)b_1^2$,
%$\lambda_4>(a_1{+}a_2)b_1^2>\lambda_2$,
%otherwise if $(a_4 b_2^2(1{+}c^2)>(a_1{+}a_2)b_1^2$,
%$\lambda_4>(a_4 b_2^2(1{+}c^2)>(a_1{+}a_2)b_1^2>\lambda_2$.
In the case of non-zero external field $h \neq 0$, one eigenvalue of the transfer matrix still equals to zero and additional three eigenvalues
are given by the roots of cubic equation
$\lambda^3+A\lambda^2+B\lambda+C=0$, where
\onecolumngrid
\begin{eqnarray}
\label{cubic_eq}
%&&\lambda^3+A\lambda^2+B\lambda+C=0,
%\nonumber\\
&& A={-}a_1^- (b_1^-)^2 {-} a_1^+ (b_1^+)^2 {-} a_4b_2^2(1+c^2),
\nonumber\\
&& B=\Big[(a_1^- (b_1^-)^2 {+} a_1^+ (b_1^+)^2)a_4 {-}(a_3^-)^2(b_1^+)^2 {-}(a_3^+)^2(b_1^-)^2 \Big]b_2^2(1+c^2)
%\nonumber\\&&
+(a_1^-a_1^+-a_2^2)(b_1^-)^2(b_1^+)^2,
\nonumber\\
&& C= \Big[-(a_1^-a_1^+ - a_2^2)a_4
+ a_1^-(a_3^+)^2 + a_1^+(a_3^-)^2 - 2a_2a_3^- a_3^+ \Big]
(b_1^-)^2(b_1^+)^2b_2^2(1+c^2).
\nonumber
\end{eqnarray}
\twocolumngrid
The roots can be calculated using trigonometric solution of cubic equation (see e. g. \cite{korn}):
\begin{eqnarray}
\label{cubic_sol}
&&\lambda_n=-2\sqrt{p}\cos\left(\frac{\phi}{3}+\frac{2\pi n}{3}\right), \quad (n=0,1,2)
%\nonumber
\\
&&p=\left(\frac{A}{3}\right)^2-\frac{B}{3}, \; \; \; \;
q=-\left(\frac{A}{3}\right)^3+\frac{AB}{6}-\frac{C}{2},
\nonumber\\
&&\cos \phi =\frac{q}{\sqrt{p^3}}.
\nonumber
\end{eqnarray}
The free energy per site in the thermodynamic limit is obtained within the transfer-matrix method \cite{baxter} as
\begin{eqnarray}
\label{free_en}
f=\lim_{N\to\infty}-\frac{1}{2N\beta}\log Z=-\frac{1}{4\beta}\log\lambda_{max},
\end{eqnarray}
where $\lambda_{max}$ denotes the maximal eigenvalue of the transfer matrix (\ref{t-matrix}).

\section{The ground state}
\label{gs_section}

Let us start by examining the ground-state properties of the spin-1/2 Ising-Heisenberg orthogonal-dimer chain.
%which is known exactly for the pure Heisenberg model only for particular ratio of parameters \cite{richter1998},
%or was obtained numerically using density matrix renormalization group method \cite{koga2000}.
To get the ground state, we have to find first the lowest-energy eigenstate of the local Hamiltonian (\ref{local_ham}).
For one-dimensional system it is then always possible to extend this state to the whole chain, which can be afterwards
proven to be the global ground state using the variational principle (see e.g. \cite{bose1992}). Using this procedure,
we have found the following six ground states:
\begin{itemize}
\item the unique {\em singlet dimer} (SD) phase
\begin{eqnarray}
\label{SD}
&&|\mbox{SD}\rangle=\prod_{i=1}^N| {\mathcal S}_{i}\rangle,
%\frac{1}{\sqrt2}(|\uparrow_{1,i}\downarrow_{2,i}\rangle -|\downarrow_{1,i}\uparrow_{2,i}\rangle),
\\
&&|{\mathcal S}_{i}\rangle
=\frac{1}{\sqrt2}(|\uparrow_{1,i}\downarrow_{2,i}\rangle -|\downarrow_{1,i}\uparrow_{2,i}\rangle)
\end{eqnarray}
with the energy $E_0=-NJ(2+\Delta)/4$,

\item the two-fold degenerate {\em modulated antiferromagnetic} (MAF) phase
\begin{eqnarray}
\label{MAF}
|\mbox{MAF}\rangle&{\!\!\!\!\!\!=}&\!\!\!\!\!\!\!{\prod_{i=1}^{N/4}}\!\!
\left\{\!\!\!
\begin{array}{l}
|{\uparrow_{1,2i}^{}}{\uparrow_{2,2i}^{}}\rangle
|{\phi_{2i+1}^{(+)}}\rangle
|{\downarrow_{1,2i+2}^{}}{\downarrow_{2,2i+2}^{}}\rangle
|{\phi_{2i+3}^{(-)}}\rangle,
\\
|{\downarrow_{1,2i}}{\downarrow_{2,2i}}\rangle
|{\phi_{2i+1}^{(-)}}\rangle
|{\uparrow_{1,2i+2}}{\uparrow_{2,2i+2}}\rangle
|{\phi_{2i+3}^{(+)}}\rangle,
\end{array}
\right.
\\
|\phi_{i}^{(+)}\rangle
&{=}&-\sin\left(\frac{\alpha}{2}\right)|\uparrow_{1,i}\downarrow_{2,i}\rangle +\cos\left(\frac{\alpha}{2}\right)|\downarrow_{1,i}\uparrow_{2,i}\rangle,
\nonumber\\
|\phi_{i}^{(-)}\rangle
&{=}&\cos\left(\frac{\alpha}{2}\right)|\uparrow_{1,i}\downarrow_{2,i}\rangle +\sin\left(\frac{\alpha}{2}\right)|\downarrow_{1,i}\uparrow_{2,i}\rangle,
\nonumber\\
\cos\alpha &{=}& \frac{2J_1}{\sqrt{J^2+4J_1^2}}
\end{eqnarray}
with the energy $E_0=-N\sqrt{J^2+4J_1^2}/4$ and the quantum reduction of the staggered magnetization of the spins residing on the horizontal Heisenberg bonds:
$\langle s^z_{1,2i+1}\rangle=\langle s^z_{2,2i+3}\rangle= -\langle s^z_{2,2i+1}\rangle=-\langle s^z_{1,2i+3}\rangle=-\cos\alpha$,

\item the two-fold degenerate {\em antiferromagnetic} (AF) phase
\begin{eqnarray}
\label{AF}
|\mbox{AF}\rangle&=&\prod_{i=1}^{N/2}
\left\{
\begin{array}{l}
|\uparrow_{1,2i}\uparrow_{2,2i}\rangle |\downarrow_{1,2i+1}\downarrow_{2,2i+1}\rangle, \\
|\downarrow_{1,2i}\downarrow_{2,2i}\rangle |\uparrow_{1,2i+1}\uparrow_{2,2i+1}\rangle
\end{array}
\right.
\end{eqnarray}
with the energy $E_0=N(\Delta J/2-J_1)/2$,

\item the two-fold degenerate {\em modulated ferrimagnetic} (MFI) phase
\begin{eqnarray}
\label{MFI}
|\mbox{MFI}\rangle&=&\prod_{i=1}^{N/4}
\left\{
\begin{array}{l}
|\uparrow_{1,2i}\uparrow_{2,2i}\rangle
|\varphi_{2i+1}^{(+)}\rangle
%|\mbox{s}_{2i+2}\rangle
|{\mathcal S}_{2i+2}\rangle
%\frac{1}{\sqrt2}(|\uparrow_{1,2i+2}\downarrow_{2,2i+2}\rangle -|\downarrow_{1,2i+2}\uparrow_{2,2i+2}\rangle)
|\varphi_{2i+3}^{(-)}\rangle,
\\
%|\mbox{s}_{2i}\rangle
|{\mathcal S}_{2i}\rangle
%\frac{1}{\sqrt2}(|\uparrow_{1,2i}\downarrow_{2,2i}\rangle -|\downarrow_{1,2i}\uparrow_{2,2i}\rangle)
|\varphi_{2i+1}^{(-)}\rangle
|\uparrow_{1,2i+2}\uparrow_{2,2i+2}\rangle
|\varphi_{2i+3}^{(+)}\rangle,
\end{array}
\right.
\\
%|\mbox{\sl S}_{2i}\rangle
%|{\mathcal S}_{2i}\rangle
%&=&\frac{1}{\sqrt2}(|\uparrow_{1,2i}\downarrow_{2,2i}\rangle -|\downarrow_{1,2i}\uparrow_{2,2i}\rangle),
%\nonumber\\
|\varphi_{i}^{(+)}\rangle
&=&-\sin\left(\frac{\alpha'}{2}\right)|\uparrow_{1,i}\downarrow_{2,i}\rangle +\cos\left(\frac{\alpha'}{2}\right)|\downarrow_{1,i}\uparrow_{2,i}\rangle,
\nonumber\\
|\varphi_{i}^{(-)}\rangle
&=&\cos\left(\frac{\alpha'}{2}\right)|\uparrow_{1,i}\downarrow_{2,i}\rangle +\sin\left(\frac{\alpha'}{2}\right)|\downarrow_{1,i}\uparrow_{2,i}\rangle,
\nonumber\\
\cos\alpha' &=& \frac{J_1}{\sqrt{J_1^2+J^2}}
\end{eqnarray}
with the energy $E_0=-N(\sqrt{J^2+J_1^2}+J(1+\Delta)/2+h)/4$ and the quantum reduction of the staggered magnetization of the spins residing on the horizontal Heisenberg bonds: $\langle s^z_{1,2i+1}\rangle=\langle s^z_{2,2i+3}\rangle= -\langle s^z_{2,2i+1}\rangle=-\langle s^z_{1,2i+3}\rangle=-\cos\alpha'$,

\item the two-fold degenerate {\em staggered bond} (SB) phase
\begin{eqnarray}
\label{SB}
|\mbox{SB}\rangle&=&\prod_{i=1}^{N/2}
\left\{
\begin{array}{l}
|\uparrow_{1,2i}\uparrow_{2,2i}\rangle
%|\mbox{s}_{2i+1}\rangle
|{\mathcal S}_{2i+1}\rangle,
%\frac{1}{\sqrt2}(|\uparrow_{1,2i+1}\downarrow_{2,2i+1}\rangle -|\downarrow_{1,2i+1}\uparrow_{2,2i+1}\rangle),
\\
%\frac{1}{\sqrt2}(|\uparrow_{1,2i}\downarrow_{2,2i}\rangle -|\downarrow_{1,2i}\uparrow_{2,2i}\rangle)
%|\mbox{s}_{2i}\rangle
|{\mathcal S}_{2i}\rangle
|\uparrow_{1,2i+1}\uparrow_{2,2i+1}\rangle
\end{array}
\right.
\end{eqnarray}
with the energy $E_0=-N(J/2+h)/2$,

\item the {\em ferromagnetic} (FM) phase
\begin{eqnarray}
\label{FM}
|\mbox{FM}\rangle&=&\prod_{i=1}^{N/2}
|\uparrow_{1,2i}\uparrow_{2,2i}\rangle |\uparrow_{1,2i+1}\uparrow_{2,2i+1}\rangle
\end{eqnarray}
with the energy $E_0=N(J_1/2+\Delta J/4 - h)$.
\end{itemize}

\begin{figure}[t]
 \begin{center}
   \epsfig{file=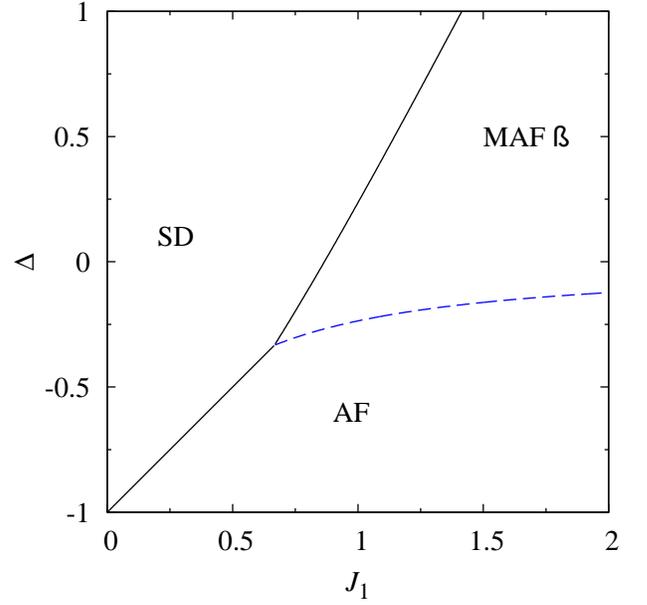, width=0.45\textwidth}\\
 \end{center}
\caption{(Color online) Ground-state phase diagram of the spin-1/2 Ising-Heisenberg orthogonal-dimer chain (\ref{gen_ham1}) in a zero magnetic field for $J=1$.
The broken line indicates a phase boundary with the macroscopic degeneracy $2^{N/2}$.}
\label{gs_pd1}
\end{figure}

The zero-field ground-state phase diagram shown in Fig.~\ref{gs_pd1} contains just three different ground states SD, MAF and AF.
Obviously, the SD phase becomes the ground state in the particular case of the antiferromagnetic Heisenberg coupling and sufficiently weak
Ising inter-dimer interaction. When the Ising inter-dimer interaction $J_1$ strengthens, the MAF phase is favored with a peculiar quantum
antiferromagnetic order of the spins from the horizontal Heisenberg bonds accompanied with the alternating character of the fully polarized triplets
on the vertical Heisenberg bonds (\ref{MAF}). Finally, the fully polarized triplet states on the vertical and horizontal Heisenberg dimers become favorable for the special case of ferromagnetic intra-dimer $zz$ coupling ($\Delta<0$), whereas the nearest-neighboring vertical and horizontal Heisenberg bonds are polarized
in opposite direction. The boundary between the relevant phases can be readily calculated by comparing the ground-state energies $E_0$ of individual phases:
\begin{itemize}
\item SD-AF: $\Delta=2J_1/J - 1$,
\item SD-MAF: $\Delta=-2+\sqrt{1+4(J_1/J)^2}$,
\item MAF-AF: $\Delta=2J_1/J-\sqrt{1+4(J_1/J)^2}$.
\end{itemize}
It is noteworthy that all curves meet at one triple point with the coordinates $J_1/J=2/3$, $\Delta=-1/3$. As it will be shown below, the ground-state boundary between MAF and AF has macroscopic degeneracy ${\mathcal W}=2^{N/2}$.

\begin{figure*}[t]
 \begin{center}
   \epsfig{file=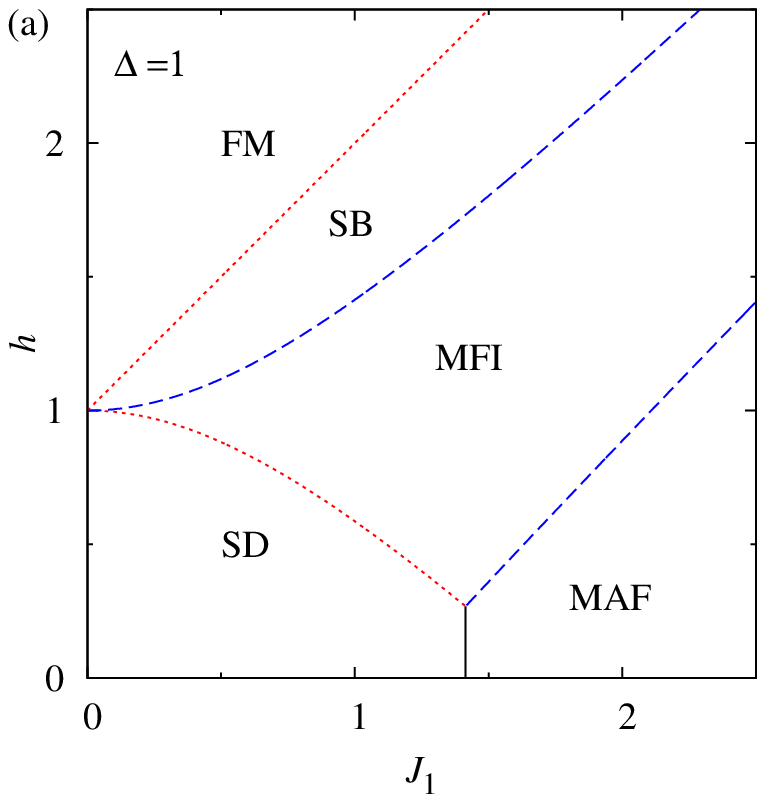, width=0.45\textwidth}
   \epsfig{file=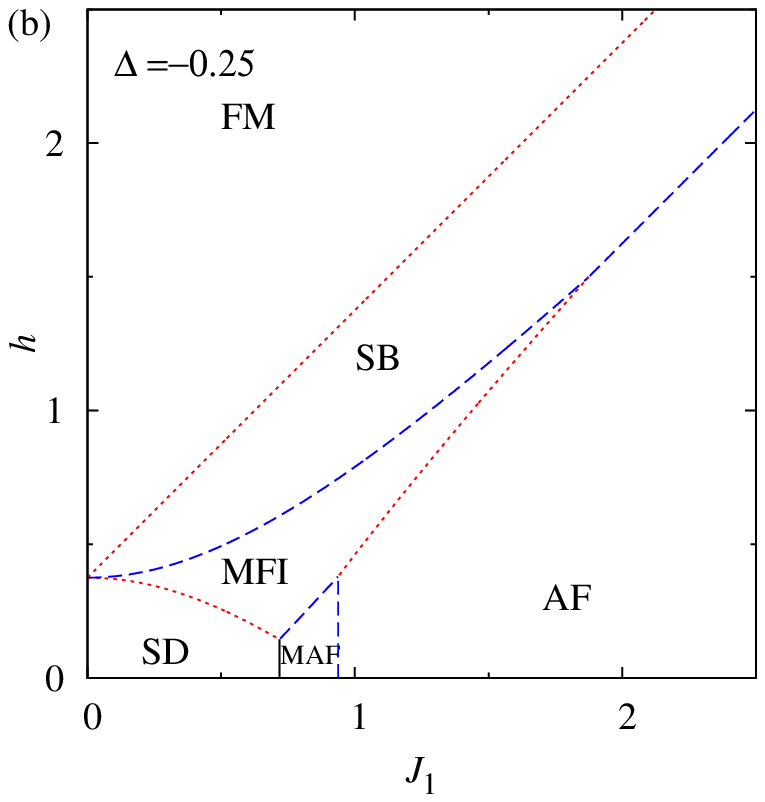, width=0.45\textwidth}
   \epsfig{file=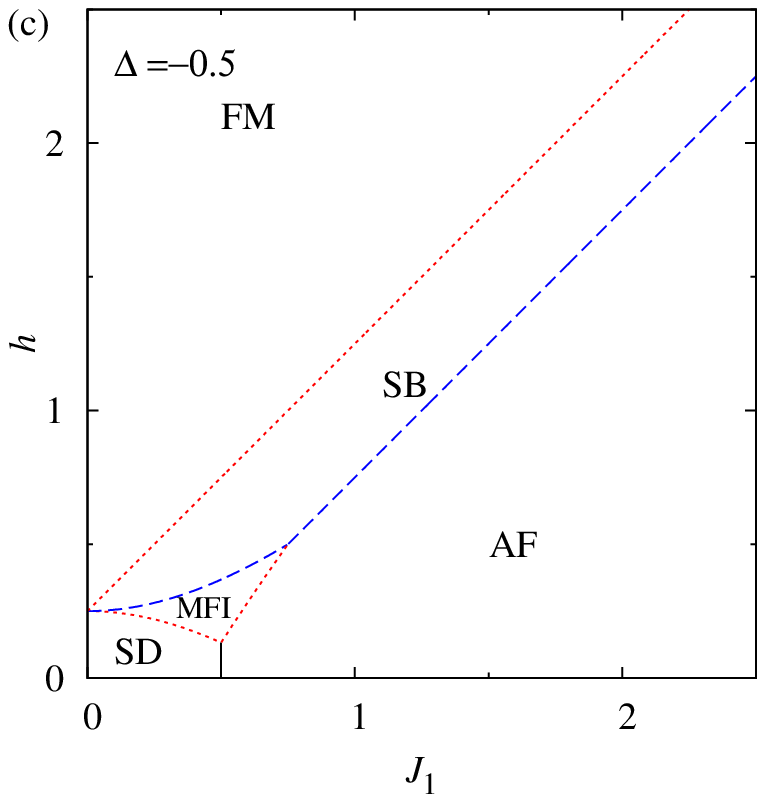, width=0.45\textwidth}
   \epsfig{file=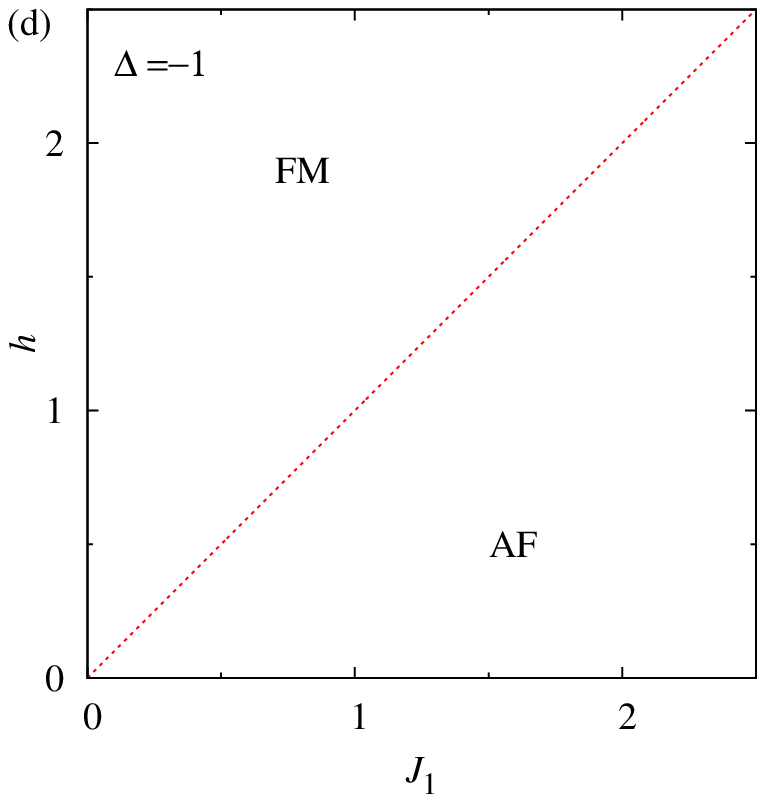, width=0.45\textwidth}
\end{center}
%\vspace*{-30pt}
\caption{(Color online) Ground-state phase diagram of the spin-1/2 Ising-Heisenberg orthogonal-dimer chain for $J=1$, $\Delta=1,-0.25,-0.5,-1$.
Broken lines denote the ground state with the macroscopic degeneracy of monomer covering on a chain,
dotted lines denote
the macroscopic degeneracy of dimer covering on a chain (see the text below).}
\label{gs_pd2}
\end{figure*}

The effect of non-zero magnetic field on the overall ground-state phase diagram is shown
in Fig.~\ref{gs_pd2} for several values of the exchange anisotropy $\Delta$.
Consider first the particular case of antiferromagnetic Heisenberg intra-dimer interaction with $\Delta=1$.
Under this condition, the zero-field ground state can be either SD or MAF phase
depending on a mutual competition between the Heisenberg intra-dimer and Ising inter-dimer interactions $J$
and $J_1$. However, the magnetic field of moderate strength destroys both SD and MAF states
due to energetic stabilization of the MFI phase,
which is characterized by the alternating character of the non-magnetic singlets
and fully polarized triplets on the vertical Heisenberg bonds. Consequently,
this latter ground state manifests itself in a magnetization process
as the intermediate one-quarter plateau, because there is no contribution to the total magnetization
from the spins on the horizontal Heisenberg bonds displaying a quantum antiferromagnetic order.
A further increase in the external magnetic field leads to a presence of the SB phase,
in which the non-magnetic singlets and the fully polarized triplets regularly alternate
on the horizontal and vertical Heisenberg bonds. As a result, the SB phase will cause
a presence of the additional plateau at one-half of the saturation magnetization.
Finally, an extremely strong magnetic field naturally flips all spins to the external-field direction,
which results in the fully polarized FM phase. To summarize our findings for the magnetization process,
the zero-temperature magnetization curve displays two different intermediate plateaux at one-quarter
and one-half of the saturation magnetization, which provide a macroscopic manifestation of two striking MFI
and SB phases of purely quantum character. It should be noted that this picture does not qualitatively change
for a more general case of the antiferromagnetic Heisenberg intra-dimer interactions with $\Delta>0$.

On the other hand, several topologies of the ground-state phase diagram are possible for the special case $\Delta<0$ corresponding to the ferromagnetic Heisenberg intra-dimer interaction. If $-\frac{1}{3}<\Delta<0$, the new AF phase can be detected in a low-field region of the ground-state phase diagram for sufficiently high values of the Ising inter-dimer interaction $J_1$.
As a consequence, for sufficiently strong inter-dimer interactions
the model exhibits a direct field-induced transition from the AF phase
towards the one-half plateau SB phase omitting the one-quarter plateau MFI phase.
If the Ising inter-dimer interaction $J_1$ is of moderate strength or sufficiently weak, the effect of external magnetic field is quite similar to the case discussed previously. The increasing magnetic field changes at the first transition field the ground state SD, MAF or AF to the one-quarter plateau MFI phase, then at the second transition field the MFI phase to the one-half plateau SB phase and finally, the SB phase at the saturation field to the fully polarized FM phase. Accordingly, the one-quarter and one-half fractional plateaux are still present in the relevant zero-temperature magnetization curves. Next, the vertical stripe corresponding to the MAF phase completely disappears from the ground-state phase diagram for $-1<\Delta<-\frac{1}{3}$. Finally, the ground-state phase diagram becomes quite simple for $\Delta\leq-1$ when the ferromagnetic Heisenberg intra-dimer interaction of the easy-axis type supports the polarized triplet states only. The ground state in an absence of the magnetic field is created by the AF phase and this order breaks at the saturation field to the FM phase. It is quite clear from the above argument that both intermediate fractional plateaux at one-quarter and one-half of the saturation magnetization vanish from the zero-temperature magnetization curve.

For completeness, let us provide the expressions for boundaries between individual phases, which have been obtained
by comparing the ground-state energies:
\begin{itemize}
\item SD-MFI: $h=-\sqrt{J^2+J_1^2}+\frac{J(3+\Delta)}{2}$, for $\Delta>-1$,
      macroscopic degeneracy ${\mathcal W}=(\frac{1+\sqrt{5}}{2})^{N/2}$ in the limit $N\to\infty$;
\item MAF-MFI: $h=\sqrt{J^2+4J_1^2}-\sqrt{J^2+J_1^2}-\frac{J(1+\Delta)}{2}$, for $\Delta>-\frac{1}{3}$,
%      macroscopic degeneracy
      ${\mathcal W}=2^{N/4}$;
\item MFI-SB: $h=\sqrt{J^2+J_1^2}-\frac{J(1-\Delta)}{2}$,
%      macroscopic degeneracy
      ${\mathcal W}=2^{N/2}$;
\item SB-FM: $h=\frac{J(1+\Delta)}{2} + J_1$,
%      macroscopic degeneracy
      ${\mathcal W}=(\frac{1+\sqrt{5}}{2})^{N}$;
\item AF-MFI: $h=-\sqrt{J^2+J_1^2}+2J_1-\frac{J(1+3\Delta)}{2}$, for $-1<\Delta<0$,
%      macroscopic degeneracy
      ${\mathcal W}=(\frac{1+\sqrt{5}}{2})^{N/2}$;
\item AF-SB: $h=-\frac{J(1+\Delta)}{2}+J_1$, for $-1<\Delta<0$,
%      macroscopic degeneracy
      ${\mathcal W}=2^{N/2}$;
\item AF-FM: $h=J_1$, for $\Delta<-1$,
%      macroscopic degeneracy
      ${\mathcal W}=(\frac{1+\sqrt{5}}{2})^{N}$.
\end{itemize}
It is worthwhile to remark that the ground state at some phase boundaries is highly degenerate.
To get the degeneracy, we can apply the notion of counting the dimer coverings on a chain
as used in Ref.~[\onlinecite{richter2005}]. Consider for instance the phase boundary between the SB and FM phases.
The FM phase can be represented as a one-dimensional lattice of the gas where all sites are empty
[$\dots\circ\circ\circ\circ\dots$], while the SB phase corresponds to the configuration where each second site is occupied
[$\dots\bullet\circ\bullet\circ\dots$].
Here, the magnetized state of the Heisenberg dimer is set to be an empty state, while the non-magnetic
singlet state of the Heisenberg dimer is represented by a filled state.
As a result, the ground state at the relevant phase boundary between these two phases can be constructed
from all configurations of particles with the following restriction: particles cannot occupy neighboring sites, i.e.
there is an infinitely strong nearest-neighbor repulsion between particles.
Such a system can be reformulated as a dimer problem, and the calculated degeneracy in the thermodynamic limit
is given by $[(1+\sqrt{5})/2]^N$  \cite{richter2005}. It is noteworthy that the same value of the degeneracy is obtained at the boundary between the FM and AF phases if the $S_i^z=-1$ ($\downarrow$) bond state is treated as an occupied site in latter case.
Similar representation can be constructed also at the SD-MFI and AF-MFI boundaries.
The only difference is that only vertical Heisenberg bonds are involved into the lattice-gas picture.
Therefore, the degeneracy is $[(1+\sqrt{5})/2]^{N/2}$.

All other boundaries have the degeneracy of monomers on a chain of size $N/2$ or $N/4$ or of the free Ising spins.
For instance, let us consider all possible configurations at the boundary between the MAF and MFI states. The MAF phase is consistent with the antiferromagnetic order of the vertical Heisenberg bonds and can be schematically presented as [$\dots\uparrow\downarrow\uparrow\downarrow\dots$]. On the contrary, the MFI phase shows on the vertical Heisenberg dimers a regular alternation of the fully polarized triplet and non-magnetic singlet bonds [$\dots\uparrow{0}\uparrow{0}\dots$]. Here, $\uparrow$($\downarrow$) denotes the $S_{2i}^z=\pm 1$ state and ${0}$ refers to the singlet state. Both phases are two-fold degenerate. In this particular case, the bond state on odd vertical Heisenberg bonds coincides for both phases, while two different states (either polarized or singlet) are available for any even vertical Heisenberg bond.
Two degrees of freedom of any even vertical Heisenberg bond can be then identified as the fictitious Ising spin $\frac{1}{2}$ and therefore,
the ground state at the boundary between the MAF and MFI phases has the macroscopic degeneracy $2^{N/4}$. The special case of the AF-SB boundary has the same picture as the MAF-MFI boundary. The only difference is that the discussion concerns not only vertical, but all the bonds. Thus, the macroscopic degeneracy at the AF-SB phase boundary is proportional to $2^{N/2}$.

The MAF and AF phases also have classical orderings (antiferromagnetic and ferromagnetic) for the vertical bonds, i.e. [$\dots\uparrow\downarrow\uparrow\downarrow\dots$] and
[$\dots\uparrow\uparrow\uparrow\uparrow\dots$], [$\dots\downarrow\downarrow\downarrow\downarrow\dots$] correspondingly. At the ground-state boundary the energy of any random ordering of vertical bond configurations will be the same, which means that the ground-state degeneracy equals to $2^{N/2}$ for this particular case. The similar case is the phase boundary between the MFI and SB phases. The only difference with respect to the previous case is that $\downarrow$ on the vertical bonds must be changed to the singlet state ${0}$.
The rest of analysis is analogous and gives the degeneracy $2^{N/2}$.

Comparing the ground-state phase diagram of the spin-1/2 Ising-Heisenberg orthogonal-dimer chain with the analogous result for the spin-1/2 Heisenberg orthogonal-dimer chain obtained by means of the extensive DMRG calculations (see Fig.~9 of Ref.~\onlinecite{koga2000}), we observe that both phase diagrams contain the regions with the zero, one-quarter and one-half magnetization plateaux. The difference between phase diagrams becomes fundamentally distinct for the sufficiently strong inter-dimer coupling $J_1$, when the dimer-plaquette phase evolves in the Heisenberg model instead of the MAF phase maintained by the Ising-Heisenberg model. This discrepancy is caused by an infinitely strong anisotropy of the Ising inter-dimer interaction. In addition, the spin-1/2 Heisenberg orthogonal-dimer chain exhibits the infinite series of magnetization steps between one-quarter and one-half magnetization plateaux \cite{schulenburg2002}, as well as, the continuous change of the magnetization from the one-half plateau to the saturation magnetization. On the other hand, the spin-1/2 Ising-Heisenberg orthogonal-dimer chain displays a high degeneracy at the saturation field that is accompanied with the respective magnetization jump instead.

\section{Thermodynamics and magnetocaloric effect}
\label{thermodynamics_section}
%All the functions can be obtained by differentiation of the free energy
%(\ref{free_en}):
%entropy $s=-(\partial f)/(\partial T)$,
%specific heat $c=T(\partial s)/(\partial T)$.

\begin{figure}[t]
 \begin{center}
   \epsfig{file=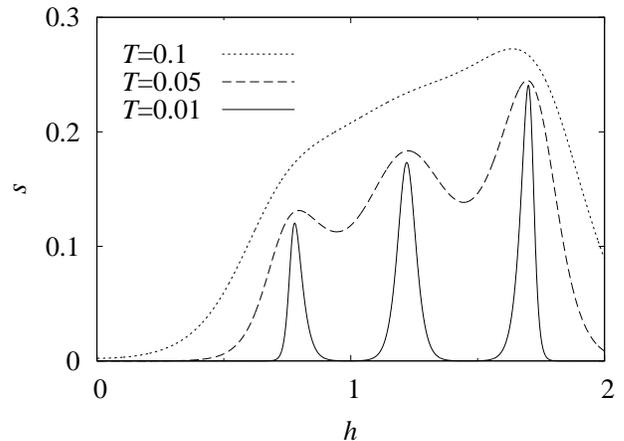, width=0.45\textwidth}
\end{center}
%\vspace*{-30pt}
\caption{The entropy per site is plotted against the magnetic field for the spin-1/2 Ising-Heisenberg orthogonal-dimer chain with
$J=1$, $\Delta=1$, $J_1=0.7$ at different temperatures $T=0.01,0.05,0.1$ (from bottom to top).}
\label{fig_entropy1}
\end{figure}

\begin{figure}[t]
 \begin{center}
   \epsfig{file=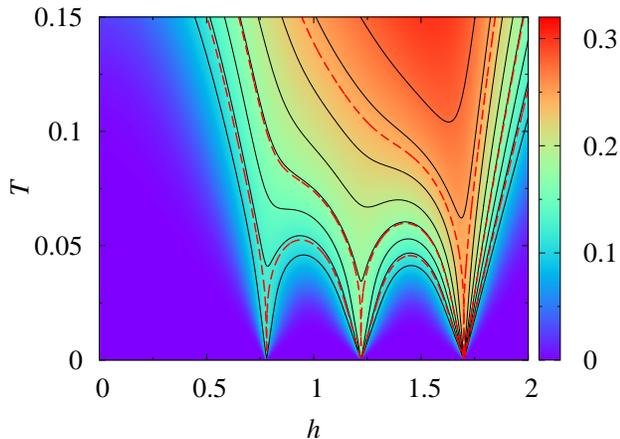, width=0.45\textwidth}
\end{center}
\caption{(Color online) A density plot of the entropy as a function of the magnetic field and temperature for the Ising-Heisenberg orthogonal-dimer chain with
$J=1$, $\Delta=1$, $J_1=0.7$. The curves with constant entropy correspond to $s=0.1, 0.125,\dots,0.3$ (solid lines) and
to $s=0.120, 0.1732, 0.2406$ (broken lines).}
\label{fig_entropy2}
\end{figure}

The macroscopic degeneracy found in the ground state may manifest itself in the low-temperature behavior of basic thermodynamic
quantities such as entropy, specific heat or magnetization. At first one should notice that the entropy can take finite values at zero temperature
whenever the ground state is macroscopically degenerate due to a phase coexistence. The entropy per site can be easily obtained using the thermodynamic relation $s=-(\partial f/\partial T)$, while the residual entropy on phase boundaries
is related to the macroscopic degeneracy of the ground-state manifold
 according to the Boltzmann's equation $s_0=\frac{1}{2N}\ln{\mathcal W}$ \cite{derzhko2004}.
Thus, the residual entropy at the ground-state boundaries between different phases can be straightforwardly calculated from the results presented in Section \ref{gs_section}. Bearing all this in mind, the residual entropy takes the value $\frac{1}{2}\ln((1+\sqrt{5})/2)\approx 0.2406$ at the SB-FM and AF-FM ground-state boundaries, $\frac{1}{4}\ln((1+\sqrt{5})/2)\approx 0.1203$ at the SD-MFI and AF-MFI boundaries, $\frac{1}{4}\ln 2 \approx 0.1733$ at the MFI-SB and AF-SB boundaries, $\frac{1}{8}\ln 2\approx 0.0866$
at the MAF-MFI boundary. As one can see from Fig.~\ref{fig_entropy1}, the field dependence of the entropy indeed shows remarkable peaks at transition fields
whose heights is in accordance with the reported values of the residual entropy (the particular case shown in Fig.~\ref{fig_entropy1} exhibits three successive field-induced transitions SD-MFI, MFI-SB and SB-FM).
It can be also clearly seen from Fig.~\ref{fig_entropy1} that even small
temperature smooths the field dependence of entropy and thus destroys its distinct profile.
Besides, it is well known that quantum frustrated spin models may exhibit an enhanced magnetocaloric effect
near the field-induced quantum critical point \cite{wolf2011,lang2013,zhitomirsky2004,ohanyan2012}.
We have therefore studied also the adiabatic demagnetization of the model under investigation, which
can be easily understood from the density plot of entropy depicted in Fig.~\ref{fig_entropy2}.
Note that the curves of constant entropy determine the change of the temperature with the magnetic field
during the adiabatic process. Since the spin-1/2 Ising-Heisenberg orthogonal-dimer chain may have
up to three critical fields accompanied with the macroscopic degeneracy of the ground state, the temperature
rapidly decreases near a critical field whenever the entropy is selected close enough to the corresponding value of the residual entropy.
This behavior may evidently promote a high adiabatic magnetocaloric rate $(\partial T/\partial h)_{S}$.
\begin{figure}[t]
 \begin{center}
   \epsfig{file=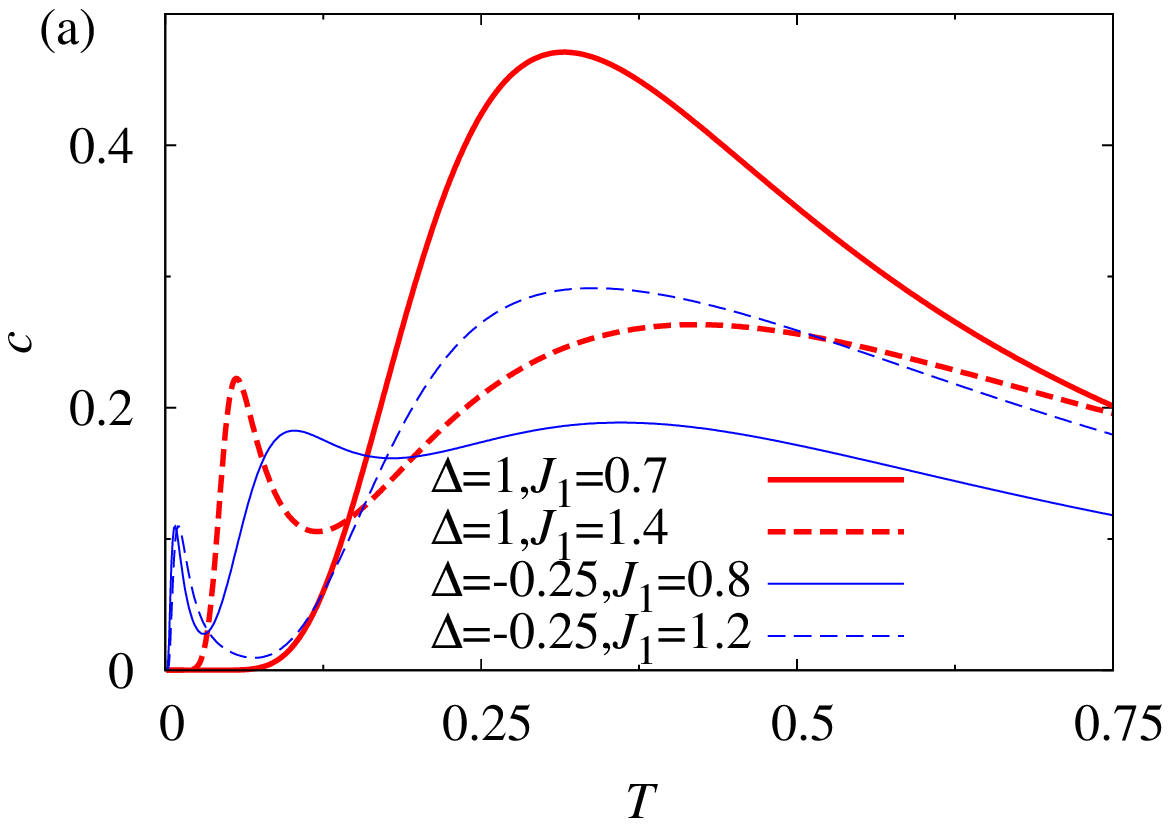, width=0.45\textwidth}
   \epsfig{file=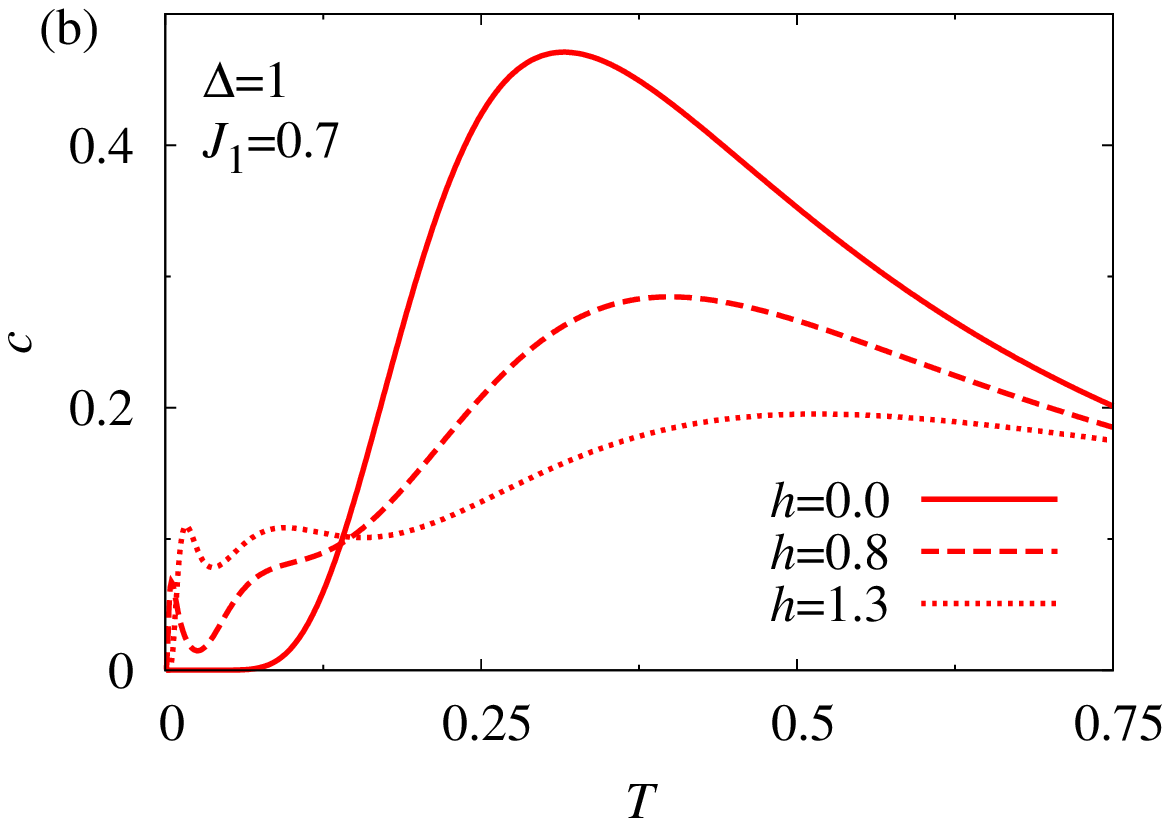, width=0.45\textwidth}
   \epsfig{file=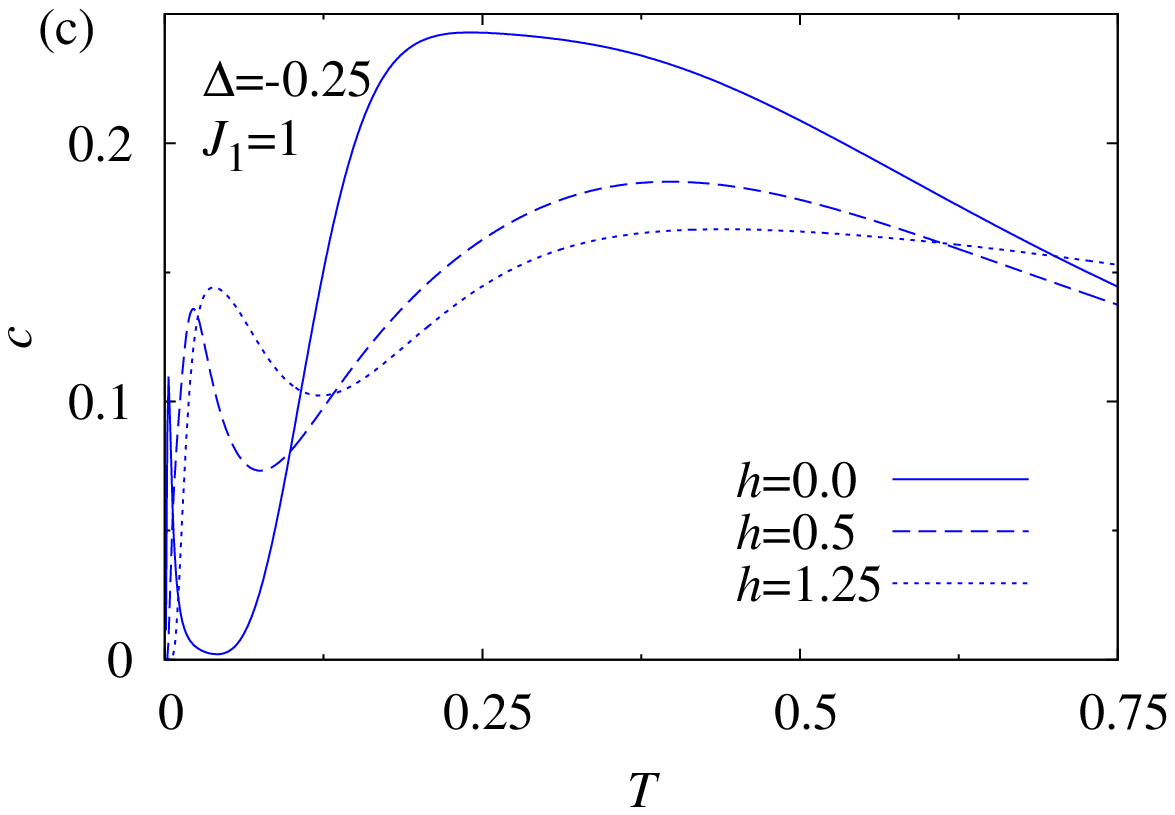, width=0.45\textwidth}
\end{center}
%\vspace*{-30pt}
\caption{(Color online) Thermal dependencies of the specific heat for:
(a) $h=0$, thick curves corresponds to $\Delta=1$, $J_1=0.7$ (solid line), $J_1=1.4$ (broken line),
thin curves corresponds to $\Delta=-0.25$,  $J_1=0.8$ (solid line), $J_1=1.2$ (broken line);
(b) $\Delta=1$, $J_1=0.7$, $h=0,0.8,1.3$;
(c) $\Delta=-0.25$, $J_1=1$, $h=0,0.5,1.25$.}
\label{fig_heat1}
\end{figure}

The obtained exact solution allows us to examine the effect of spin frustration and external field on the specific heat,
which can be obtained from the thermodynamic relation $c=T(\partial s/\partial T)$. Some typical thermal variations
of the specific heat are presented in Fig.~\ref{fig_heat1} for different values of the interactions and external magnetic field.
The temperature dependencies of zero-field specific heat are displayed in Fig.~\ref{fig_heat1}(a).
The investigated spin system is far from the degenerate ground state for the special case $\Delta=1$, $J=1$, $J_1=0.7$
and hence, the specific heat exhibits just one broad peak of Schottky type. On the other hand, the zero-field specific heat
gains an additional low-temperature peak by changing the Ising inter-dimer coupling $J_1=1.4$ sufficiently close to the SD-MAF boundary.
The set of parameters driving the investigated spin chain close to the macroscopically degenerate MAF-AF boundary
shows even more complex temperature dependence with rapidly increasing specific heat at low temperature and several round maxima.

Fig.~\ref{fig_heat1}(b) and (c) illustrate thermal variations of the specific heat when the external field is selected close to critical fields.
In Fig.~\ref{fig_heat1}(b), the applied magnetic field $h=0.8$ is sufficient to stabilize the one-quarter plateau with a quite small gap between
the ground state and first excited state. When the external field achieves the critical value, the ground state becomes macroscopically degenerate.
The specific heat then shows a sharp peak at very low temperature and quite broad nearly flat region between two peaks. This unusual dependence
indicates the existence of a
large number of states with energies quite close to the ground state energy. The specific heat near the MFI-SB border
(case of $h=1.3$ in Fig.~\ref{fig_heat1}(b)) shows even more striking temperature dependence with three peaks, whereas all three peaks are of the same order.
The specific heat for the other particular case of the ferromagnetic Heisenberg intra-dimer coupling shown in Fig.~\ref{fig_heat1}(c) has similar features. It should be nevertheless mentioned that the zero-field specific heat shows a very sharp low-temperature peak for the interaction parameters driving the investigated spin chain close to the MAF-AF ground-state boundary. The external field generally broadens this peak and shifts it towards slightly higher temperatures.

\begin{figure}[t]
 \begin{center}
   \epsfig{file=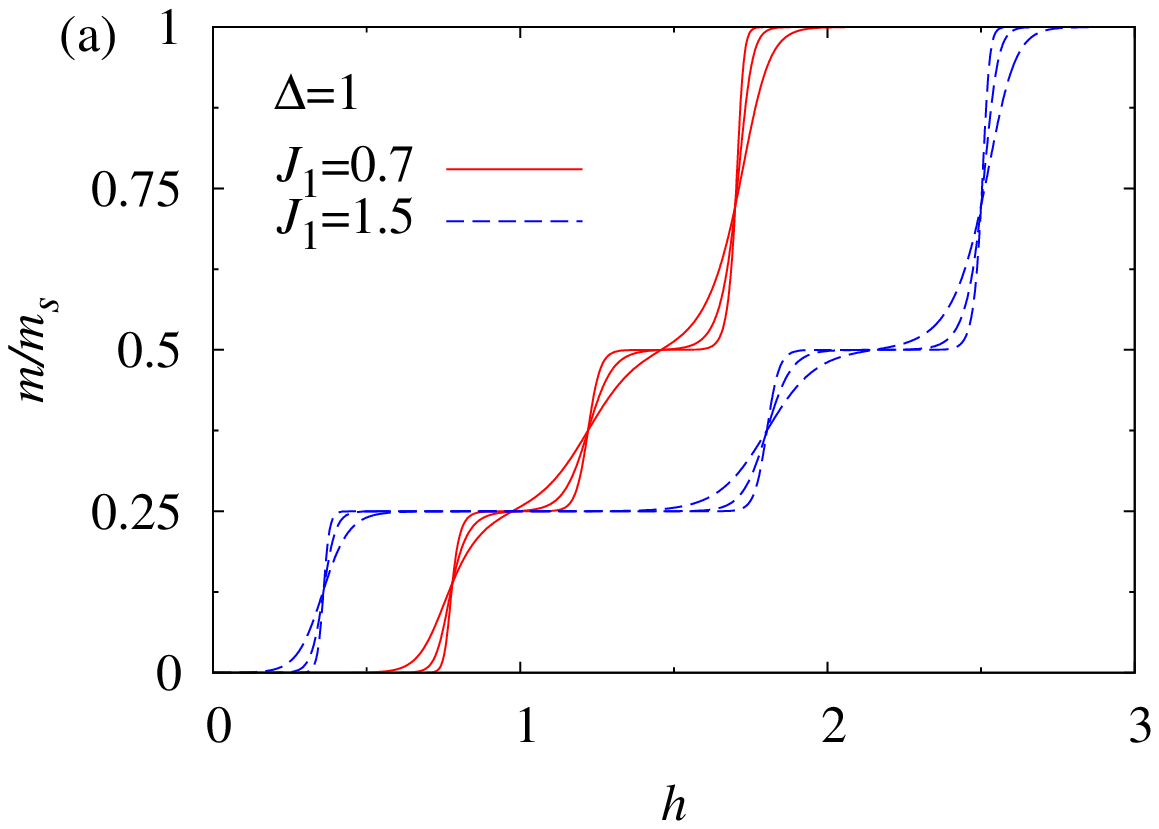, width=0.45\textwidth}
   \epsfig{file=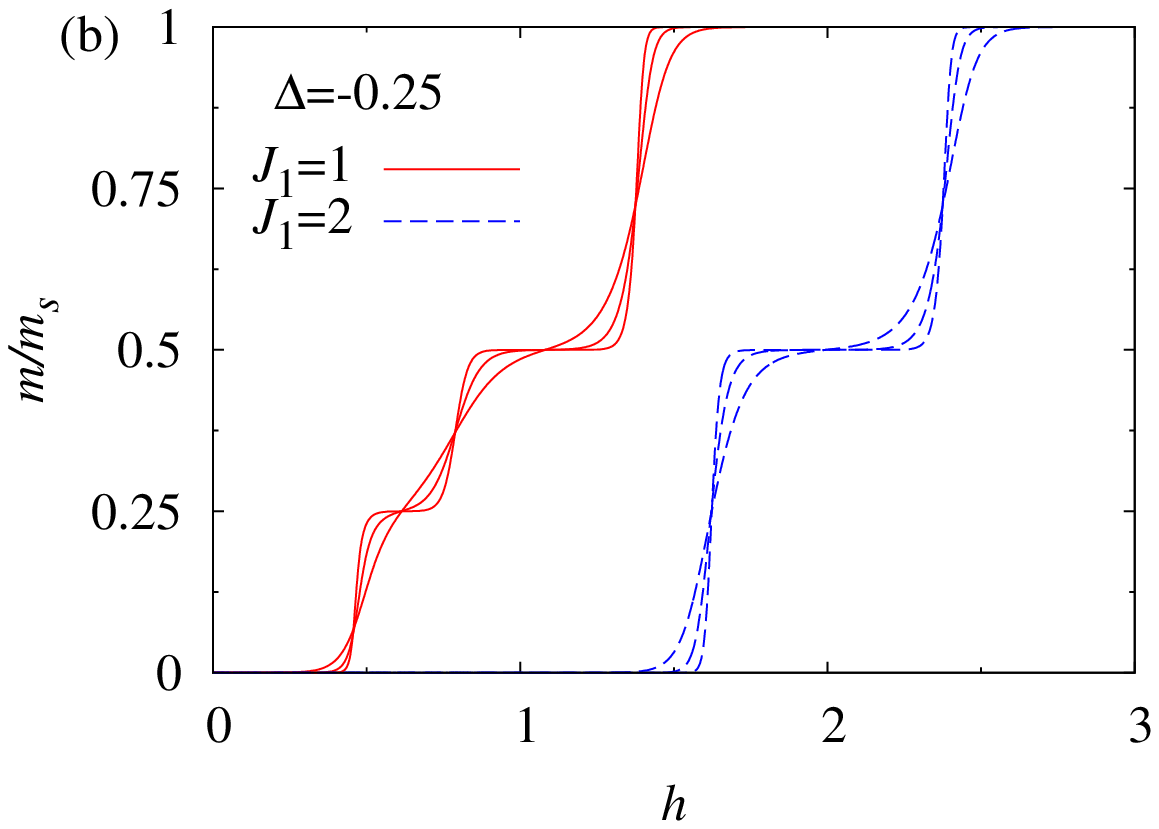, width=0.45\textwidth}
\end{center}
%\vspace*{-30pt}
\caption{(Color online) The magnetization normalized with respect to its saturation value as a function of the magnetic field for:
(a) $J=1$, $\Delta=1$, $T=0.01,0.02,0.04$ smother curve corresponds to higher temperature.
Solid lines  corresponds to $J_1=0.7$, broken lines corresponds to $J_1=1.5$;
(b) $J=1$, $\Delta=-0.25$, $T=0.01,0.02,0.04$.
Solid lines  corresponds to $J_1=1$, broken lines corresponds to $J_1=2$.}
\label{fig_mag}
\end{figure}

Let us also briefly comment on a magnetization process of the spin-1/2 Ising-Heisenberg orthogonal-dimer chain
at low temperatures. It is quite evident from the ground-state phase diagram shown in Fig.~\ref{gs_pd2} that the low-temperature
magnetization curve may contain fractional plateaux at one-quarter and one-half of the saturation magnetization. As one could expect,
the intermediate plateaux and magnetization jumps gradually become smoother as temperature increases. However, it is quite surprising
how fast a step-like magnetization curve is demolished by even very low temperature. For illustration, we present in Fig.~\ref{fig_mag}(a)
the relevant low-temperature magnetization curves for the isotropic Heisenberg intra-dimer interaction. The width of both intermediate plateaux
is nearly the same for the particular case $J_1=0.7$ and it actually turns out that even a rather small temperature $T=0.04$ makes the step-like
structure almost indistinguishable within the relevant magnetization curve. Contrary to this, the other particular case $J_1=1.5$ seems to be
more robust against thermal fluctuations when the intermediate plateaux and magnetization jumps cannot be discerned in the relevant magnetization
curve just at slightly higher temperature. Similar trends can be observed in the low-temperature magnetization curves of the investigated model
with the ferromagnetic Heisenberg intra-dimer interaction depicted in Fig.~\ref{fig_mag}(b). Under this condition, the magnetization curve may contain
either one or two intermediate plateaux depending on an interplay between the interaction parameters. In general, it could be concluded that
the rather rapid thermal smoothing can be attributed to the huge degeneracy of the ground state at critical fields.

\section{Conclusions}
\label{conclusions}

In the present work we considered the orthogonal-dimer chain with the Heisenberg intra-dimer and Ising inter-dimer interactions by means of a rigorous approach based on the transfer-matrix method. We have obtained the exact expressions for the partition function and analyzed the ground state and thermodynamic properties of the model quite rigorously. The ground-state phase diagram of the model in a magnetic field has been obtained and it was shown that
two fractional plateaux at one-quarter and one-half of the saturation magnetization are present. We have also studied the effect of the exchange anisotropy
in the Heisenberg coupling. It has been shown that the ferromagnetic $zz$ intra-dimer interaction may lead to the appearance of new phases in zero field and may substantially change the ground-state phase diagram in a non-zero magnetic field. In general, this kind of interaction leads to the vanishing of one-quarter and one-half plateaux. The ground state at the border between different phases may exhibit a high macroscopic degeneracy, which leads to the non-zero residual entropy.
We have calculated the degeneracy and the residual entropy at all boundaries using the notion of the monomer or dimer covering of a chain \cite{richter2005}.
The degenerate or nearly degenerate ground state has turned out to basically affect the low-temperature thermodynamics of
the model. We have calculated the entropy as a function of temperature and magnetic field, which evidences an enhanced magnetocaloric effect close to critical fields. The effect of spin frustration and magnetic field on temperature dependence of specific heat has been examined in detail. It has been found that the interplay of all factors may lead to the complex low-temperature behavior of the specific heat with several more or less separated maxima. The exact results
for the magnetization curves have proved that even small temperature may destroy the step-like field dependence of the magnetization.

We have also found that the Ising-Heisenberg model on the orthogonal-dimer chain exhibits some common features with the analogous pure Heisenberg model, for instance a presence of the one-quarter and one-half magnetization plateaux. The main discrepancy between both models is as follows: when the Heisenberg model shows step-like magnetization between one-quarter and one-half plateaux and a continuous change of the magnetization above the one-half plateau, the Ising-Heisenberg model cannot capture those features as it possesses macroscopic degeneracy at critical fields only and shows just two intermediate plateaux. It could be expected, however, that the treatment of the quantum $XY$ part of inter-dimer interaction within the perturbation theory for degenerate states could restore some features of the magnetization curve of the pure Heisenberg model when starting from the exactly solved Ising-Heisenberg model.

Finally, it should be mentioned that there exist an extensive series of heterobimetallic coordination polymers [Ln(hfac)$_2$(CH$_3$OH)]$_2$[Cu(dmg)(Hdmg)]$_2$ \cite{ueki2005,ueki2007,okazawa2008,okazawa2009,okazawa2011} with the magnetic structure similar to the considered model. In addition, the dysprosium-based member [Dy$_2$Cu$_2$]$_n$ of this series provides an interesting experimental realization of the spin-1/2 Ising-Heisenberg orthogonal-dimer chain owing to a strong magnetic anisotropy of Dy$^{3+}$ ions. Although a more complete description of the polymeric coordination compound [Dy$_2$Cu$_2$]$_n$ would require an analysis based on the more general (asymmetric) spin-1/2 Ising-Heisenberg orthogonal-dimer chain with four different exchange couplings, the AF ground state reported for the symmetric spin-1/2 Ising-Heisenberg orthogonal-dimer chain with just two different exchange couplings already correctly reproduces the ferrimagnetic spin arrangement observed experimentally due to the antiferromagnetic inter-dimer and ferromagnetic intra-dimer interactions \cite{ueki2007,okazawa2008,note}. Moreover, the procedure elaborated in the present work can be rather straightforwardly adopted also for a theoretical treatment of the more general (asymmetric) spin-1/2 Ising-Heisenberg orthogonal-dimer chain, which would ensure a more correct description of the heterobimetallic complex [Dy$_2$Cu$_2$]$_n$. In this direction we will continue our further efforts.

\acknowledgements
T.V. acknowledges the support of the National Scholarship Programme of the Slovak Republic. J.S. acknowledges the financial support provided under the grant VEGA 1/0234/12.


\begin{thebibliography}{99}

\bibitem{richter1998} N.B. Ivanov, J. Richter,
  Phys. Lett. A {\bf 232}, 308 (1997);
  J. Richter, N. B. Ivanov, and J. Schulenburg,
%  The antiferromagnetic spin- chain with competing dimers and plaquettes: numerical versus exact results,
  J. Phys.: Condens. Matter {\bf 10}, 3635 (1998).

\bibitem{koga2000} A. Koga, K. Okunishi, and N. Kawakami,
%  First-order quantum phase transition in the orthogonal-dimer spin chain,
  Phys. Rev. B {\bf 62}, 5558 (2000).

\bibitem{miyahara2011} Sh. Miyahara,
   Exact results in frustrated quantum magnetism.
   In: Introduction to Frustrated Magnetism, C.~Lacroix, Ph.~Mendels, F.~Mila (eds.),
  Springer series in Solid-state Science {\bf 164}, Springer-Verlag Berlin Heidelberg (2011), p.513.

%   \cite{solstsci164}, p.513.

\bibitem{miyahara2005} Sh. Miyahara, and K. Ueda,
%  Theory of the orthogonal dimer Heisenberg spin model for SrCu$_2$(BO$_3$)$_2$,
  J. Phys.: Condens. Matter {\bf 15}, R327 (2003).

\bibitem{takigawa2011} M. Takigawa and F. Mila,
  Magnetization Plateaus,
   In: Introduction to Frustrated Magnetism, C.~Lacroix, Ph.~Mendels, F.~Mila (eds.),
  Springer series in Solid-state Science {\bf 164}, Springer-Verlag Berlin Heidelberg (2011), p.241.

\bibitem{takigawa2013}
  M. Takigawa, M. Horvati\'{c}, T. Waki, S. Kr\"{a}mer, C.~Berthier, F.~L\'{e}vy-Bertrand, I.~Sheikin,
  H.~Kageyama, Y.~Ueda, and F.~Mila,
  %Incomplete Devil’s Staircase in the Magnetization Curve of SrCu$_2$(BO$_3$)$_2$,
  Phys. Rev. Lett. {\bf 110}, 067210 (2013).

\bibitem{matsuda2013}
Y. H. Matsuda, N. Abe, S. Takeyama, H. Kageyama, P. Corboz, A. Honecker, S. R. Manmana, G. R. Foltin, K. P. Schmidt, F. Mila,
 	arXiv:1308.4151 [cond-mat.str-el].

\bibitem{shastry1981} B. S. Shastry, B. Sutherland,
%  Exact ground state of a quantum mechanical antiferromaget,
  Physica B \& C {\bf 108}, 1069 (1981).

\bibitem{manmana2011} S. R. Manmana, J.-D. Picon, K. P. Schmidt, and F. Mila,
%  Unconventional magnetization plateaus in a Shastry-Sutherland spin tube,
  Europhys. Lett. {\bf 94}, 67004 (2011).

\bibitem{schulenburg2002} J. Schulenburg, and J. Richter,
%  Infinite series of magnetization plateaus in the frustrated dimer-plaquette chain,
  Phys. Rev. B {\bf 65}, 054420 (2002).

\bibitem{schulenburg2002a} J. Schulenburg, and J. Richter,
  Phys. Rev. B {\bf 66}, 134419 (2002).

\bibitem{ohanyan2012} V. Ohanyan, A. Honecker,
%   Magneto-thermal properties of the Heisenberg-Ising orthogonal-dimer chain with triangular XXZ-clusters,
     Phys. Rev. B {\bf 86}, 054412 (2012). %[ Preprint: arXiv:1203.4741v1 [cond-mat.str-el]].

\bibitem{paulinelli2013} H. G. Paulinelli, S. M. de Souza, O. Rojas,
     J. Phys.: Condens. Matter {\bf 25}, 306003 (2013).

\bibitem{heu10} W. Van den Heuvel and L.F. Chibotaru, Phys. Rev. B \textbf{82}, 174436 (2010).

\bibitem{sah12} S. Sahoo, J.P. Sutter, S. Ramasesha, J. Stat. Phys. \textbf{147}, 181 (2012).

\bibitem{str12} J. Stre\v{c}ka, M. Hagiwara, Y. Han, T. Kida, Z.~Honda, M.~Ikeda, Condens. Matter Phys. \textbf{15}, 43002 (2012).

\bibitem{han13}
Y. Han, T. Kida, M.~Ikeda, M. Hagiwara, J. Stre\v{c}ka, Z.~Honda, J. Korean Phys. Soc. \textbf{62}, 2050 (2013).

\bibitem{ueki2005}
S. Ueki, Y. Kobayashi, T. Ishida, T. Nogami, Chem. Commun., 5223 (2005).

\bibitem{ueki2007} S. Ueki, A. Okazawa, T. Ishida, T. Nogami, H. Nojiri, Polyhedron {\bf 26}, 1970 (2007).

\bibitem{okazawa2008} A. Okazawa, T. Nogami, H. Nojiri, T. Ishida, Chem. Mater. {\bf 20}, 3110 (2008)

\bibitem{okazawa2009} A. Okazawa, R. Watanabe, H.~Nojiri, T.~Nogami, T.~Ishida, Polyhedron {\bf 28}, 1808 (2009).

\bibitem{okazawa2011} A. Okazawa, K. Fujiwara, R. Watanabe, N.~Kojima, Sh.~Yoshii, H.~Nojiri, T.~Ishida,
   Polyhedron {\bf 30}, 3121 (2011).

\bibitem{rojas2012} O. Rojas, J. Stre\v{c}ka, M.L. Lyra,
    Phys. Lett. A {\bf 377}, 920 (2013).

\bibitem{baxter} R.J. Baxter, Exactly Solved Models in Statistical Mechanics (Academic, New York, 1982).

\bibitem{korn}  G.A. Korn, Th.M. Korn,
    Mathematical Handbook for Scientists and Engineers: Definitions, Theorems, and Formulas for Reference and Review
    (Dover Publications, 2000).

\bibitem{bose1992} I. Bose, Phys. Rev. B {\bf 45}, 13072  (1992).

\bibitem{derzhko2004} O. Derzhko, J. Richter,
   Phys. Rev. B 7{\bf 70}, 104415 (2004).

\bibitem{richter2005} J. Richter,
%Localized-magnon states in strongly frustrated quantum spin lattices,
Fizika Nizkikh Temperatur (Kharkiv) {\bf 31}, 918 (2005) [Low Temperature Physics {\bf 31}, 695 (2005)]

\bibitem{zhitomirsky2004} M.E. Zhitomirsky, A. Honecker,
   J. Stat. Mech.: Theor. Exp. P07012 (2004).

\bibitem{wolf2011} B. Wolf, Y. Tsui, D. Jaiswal-Nagar, U. Tutsch, A. Honecker, K. Removi\'c-Langer, G. Hofmann, A. Prokofiev,
    W. Assmus, G. Donath, and M. Lang,
    Proceedings of the National Academy of Sciences  {\bf 108},
    6862 (2011).

\bibitem{lang2013} M. Lang, B. Wolf, A. Honecker, L. Balents, U. Tutsch, P.~T. Cong, G. Hofmann,
    N. Kr{\"u}ger, F. Ritter, W. Assmus, A. Prokofiev,
    Physica Status Solidi B Basic Research, {\bf 250}, 457 (2013).

\bibitem{note}
Note that we have assumed in the present work equal $g$-factors for the horizontal and vertical spin-1/2 dimers, what consequently leads to the AF ground state with zero total magnetization in a parameter space with the antiferromagnetic inter-dimer and ferromagnetic intra-dimer interactions. Even though Dy$^{3+}$ and Cu$^{2+}$ magnetic ions can be treated as the spin-1/2 entities at low enough temperatures, the difference in the relevant $g$-factors ($g_{\rm Dy} \approx 20$ vs. $g_{\rm Cu} \approx 2$) is responsible in [Dy$_2$Cu$_2$]$_n$ for the ferrimagnetic ground state with a non-zero total magnetization that however exactly coincides with the spin arrangement reported for the AF ground state.

\end{thebibliography}
\end{document}